\documentclass[
    aip,
    amsmath,amssymb,
    reprint,%
]{revtex4-2}

\usepackage{graphicx}
\usepackage{dcolumn}
\usepackage{bm}
\usepackage[english]{babel}
\usepackage{natmove}

\usepackage[utf8]{inputenc}
\usepackage[T1]{fontenc}
\usepackage{mathptmx}
\usepackage{soul}
\usepackage{tikz}
\usepackage{xcolor}
\usepackage{amsmath}
\usepackage{hyperref}

\usepackage[normalem]{ulem} 
\usepackage{subcaption}

\newcommand\rout{\bgroup\markoverwith
{\textcolor{red}{\rule[.5ex]{2pt}{2pt}}}\ULon}

\newcommand*{\bfit}[1]{\textbf{\textit{#1}}}
\newcommand*{\epp}[0]{ESPResSo++}

\begin{document}
\title{Structure and dynamics of ionic liquids under shear flow}
\author{Abbas Gholami}
\thanks{Equally-contributing first author}
\email{gholamia@mpip-mainz.mpg.de}
\affiliation{Max Planck Institute for Polymer Research, 55128 Mainz, Germany}

\author{Sebastian Kloth}
\thanks{Equally-contributing first author}
\affiliation{Institute of Condensed Matter Physics, Technische Universit\"at Darmstadt, 64289 Darmstadt, Germany}

\author{Zhen-Hao Xu}
\affiliation{Center for Data Processing, Johannes Gutenberg-Universität Mainz, 55128 Mainz, Germany}

\author{Kurt Kremer}
\affiliation{Max Planck Institute for Polymer Research, 55128 Mainz, Germany}

\author{Michael Vogel}
\affiliation{Institute of Condensed Matter Physics, Technische Universit\"at Darmstadt, 64289 Darmstadt, Germany}

\author{Torsten Stuehn}
\affiliation{Max Planck Institute for Polymer Research, 55128 Mainz, Germany}

\author{Joseph F. Rudzinski}
\affiliation{Physics Department and CSMB Adlershof, Humboldt-Universität zu Berlin, 12489 Berlin, Germany}
\affiliation{Max Planck Institute for Polymer Research, 55128 Mainz, Germany}

\date{\today}

\begin{abstract}
    We investigate the intrinsic behavior of ionic liquids under shear flow, using a coarse-grained model of [C$_4$mim]$^+$ [PF$_6$]$^-$ as a prototypical example. The importance of long-ranged electrostatics is assessed as a function of shear rate by comparing Ewald and reaction field treatments. An appropriate comparison is achieved through the implementation of the proper Lees-Edwards boundary conditions within the \epp~simulation software. 
    Our results demonstrate that while structural properties are relatively insensitive to the electrostatic treatment, the more accurate treatment via the Ewald approach is essential for studies of dynamics, in particular, at lower shear rates. Furthermore, we identify a critical shear rate beyond which structural and dynamical properties begin to deviate from equilibrium behavior, while remaining largely unchanged below this threshold. Finally, we demonstrate that the dynamic heterogeneity of the liquid decreases as a function of increasing shear rate, which can be primarily explained by the faster dynamics induced by the shear flow. These results hold relevance for investigations of process-dependent properties of ionic-liquid-based materials.

\end{abstract}
\maketitle
\section{Introduction}\label{sec:introduction}

 Ionic liquids (ILs) show immense promise for a variety of applications, for example, for sustainable energy storage devices (e.g., for lithium batteries and fuel cells~\cite{Watanabe_CR_17}) and for improved lubrication technologies~\cite{Somers:2013}. These applications aim for more energy-efficient or safer materials by utilizing the unique physicochemical properties of ILs, e.g., low vapor pressure, high thermal stability, non-flammability, and facile ion transport. By selecting ``base'' IL cations and anions from the wide variety of available systems and through molecular variations of these ions via, e.g., changes in side chain length or polymerization, the properties of the resulting materials can be precisely tuned. However, the relationship between molecular features and material properties is extremely difficult to characterize.  Molecular simulations have the potential to provide significant insight into this relationship, assisting in the design of next-generation technologies. Coarse-grained (CG) models, in particular, provide access to length and time scales necessary for the investigation of these complex materials.

There is already a large collection of previous work focused on investigating and assessing CG model behavior for ILs at equilibrium, while little is known about these models' behavior under non-equilibrium conditions. More explicitly, previous works have:
\begin{enumerate}
\item{Explicitly compared CG ionic liquid structure and dynamics with all-atom (AA) models~\cite{Vogel_JCP_15_PH, Vogel_CPC_17},}
\item{Built CG force fields for imidazolium-based ILs directly from a reference AA model (i.e., using ``bottom-up'' techniques) to reproduce certain structural and thermodynamic properties~\cite{doi:10.1021/ct800548t, doi:10.1021/acs.jctc.7b01293, kloth_coarse-grained_2021}. These studies investigated transferability over chain length and thermodynamic state, while also pursuing large-scale simulations to explore emergent properties, e.g., nanoscale segregation upon cooling.}  

\item{Developed CG IL models based on the ``top-down'' Martini force field, while aiming to also reproduce certain structural properties of the system~\cite{D0GC01823F, CRESPO2020324}. These studies investigated the effect of temperature, alkyl chain length, IL concentration, and cation chemistry on the phase behavior of the system.}
\item{Characterized the dominant dynamic modes of ILs and assessed the effect of electrostatic interactions and temperature on these modes, while also characterizing the impact on the system's overall dynamic heterogeneity~\cite{Mueller-Plathe_PCCP_10, Vogel_CPC_17, Vogel_JCP_19_TP}.}
\item{Systematically compared the performance of various CG strategies in preserving the relations between distinct dynamical modes of ILs~\cite{Rudzinski_2021}.} 
\end{enumerate}

The difficulty in connecting molecular features with material properties is due in no small part to ion transport properties, which are complicated by the non-Newtonian nature of ILs and may strongly depend on the coupling (or lack thereof) between ion diffusion and characteristic motions of other molecular components dissolved in IL materials~\cite{Shakeel_ACS_19}. Moreover, the morphology of ILs, in particular, for composite materials is often intimately connected to the processing conditions (e.g., via solvent casting). 
Thus, although the above-mentioned studies have already provided important insight into the properties of CG IL models, investigations of these models under non-equilibrium conditions are essential.
 
As a significant step in this direction, the present study presents an investigation of CG IL structure and dynamics under shear flow. 
The shear is applied by ensuring the appropriate Lees-Edwards boundary conditions~\cite{LE, 10.1063/1.3537974, 10.1122/8.0000365} (LEbc) while also incorporating accurate treatment of long-ranged electrostatics with the Ewald approach (EW)~\cite{ESPP_LEBC_2021}.
The boundary conditions enforce a velocity gradient between periodic images of the simulation box, enabling the flow of particles within different layers of the system. Despite its importance for avoiding artifacts in shear flow simulations, there are relatively few simulation packages that explicitly implement LEbc. 
Instead, many implementations within simulation packages, e.g., the SLLOD algorithm~\cite{10.1063/1.1819869, Cho2017} in LAMMPS~\cite{THOMPSON2022108171, Ness2023}, or the SFF approach (Shear Flow Field) in NAMD~\cite{Kang2012-jm, 10.1063/5.0014475}, mimic the conditions via box deformation.
Moreover, the option for Ewald-based long-range electrostatics is often not enabled due to the non-trivial combination with the boundary conditions, especially with proper LEbc implementations. For example, the LEbc is used to implement shear in the ESPResSo MD package without enabling long-range interactions or Ewald-based electrostatics\cite{10.1063/5.0055396}. In one case that we found, an implementation of LEbc with long-range electrostatics does appear to have been used, but only by using a pre-released branch version of Gromacs~\cite{Boushehri2024, GROMACS_MR3141}.  
Our implementation of LEbc in combination with EW electrostatics is available within the main branch of the \epp~simulation software~\cite{GUZMAN201966, HALVERSON20131129, ESPP_LEBC_2021}, making the technique more visible and accessible to the community.

As in many previous studies, we consider [C$_4$mim]$^+$ [PF$_6$]$^-$ as a prototypical IL. We assess various structural and dynamical properties, including dynamic heterogeneity, of this IL as a function of increasing shear rate. In particular, we provide a detailed investigation of the effect of long-ranged electrostatics on these properties by comparing results from the EW and reaction field (RF) methods.
This foundational investigation opens the door for future assessment of the dynamical consistency of such CG models, with respect to atomistic simulations and experiments.
Thus, the results hold relevance for the utilization of CG models for investigating the process-dependent properties of ionic-liquid-based materials.

\section{Methods}\label{sec:methods}

\subsection{Non-equilibrium Molecular Dynamics for a Shear Flow Simulation}\label{subsec:thermostat}

For driving a shear flow in ionic liquid systems with non-equilibrium molecular dynamics, we employed the SLLOD equations of motion~\cite{SLLOD_Petravic1998,SLLOD_Todd2007,SLLOD_ToddDaivis2017} and applied the Lees-Edwards boundary condition (LEbc)~\cite{LEBC_Lees1972}.
The SLLOD equations of motion are given as
\begin{subequations}
    \begin{align}
        \dot{\mathbf{r}}_i&=\frac{\mathbf{p}_i}{m_i}+\mathbf{r}_i\cdot\nabla \mathbf{v} ,\\
        \dot{\mathbf{p}}_i&=\mathbf{F}_i-\mathbf{p}_i\cdot\nabla \mathbf{v} .
    \end{align}
    \label{eq:sllod}
\end{subequations}
In Eq.~\ref{eq:sllod}, $\mathbf{r}_i$ and $\mathbf{p}_i$ are the position and momentum of atom (or bead) $i$, respectively. $\mathbf{F}_i$ is all forces acted on $i$ due to potential energy functions, constraints etc., and $\nabla \mathbf{v}$ stands for the gradient tensor to the stream velocity.
For the planar shear flow, in this work, the stream velocity applies on the $x$-direction and the gradient and vorticity directions are assigned respectively to the $z$- and $y$-directions.
Hence, $\nabla \mathbf{v}$ is simply given as
\begin{equation}
    \begin{aligned}
        \nabla \mathbf{v}=\begin{pmatrix}
                              0            & 0 & 0 \\
                              0            & 0 & 0 \\
                              \dot{\gamma} & 0 & 0 
        \end{pmatrix} ,
    \end{aligned}
    \label{eq:vel_grad}
\end{equation}
where $\dot{\gamma}$ is the rate of a planar shear flow. We can express the expected form of the streaming velocity under steady shear as
\begin{equation}
    \begin{aligned}
        \dot{\mathbf{r}}_i=\frac{\dot{\mathbf{p}}_i}{m_i}+\dot{\gamma}\cdot(z-\dfrac{L_z}{2}) ,
    \end{aligned}
    \label{eq:shearspeed}
\end{equation}
where $L_z$ is one of the dimensions in an orthogonal simulation box ($L=\left[L_x,L_y,L_z\right]$). It is noteworthy to mention that the velocity term in Eq.~\ref{eq:shearspeed} is not externally enforced but rather reflects the expected steady-state streaming velocity profile that arises under the action of the SLLOD dynamics and LEbc. 
With the presence of LEbc, the shear contribution is taken into account for wrapping both coordinates and velocities when a particle moves out of the simulation box through the top and bottom boundaries (in $xy-$plane).
More details on the implementation of LEbc in \epp~can be found in previous work~\cite{ESPP_LEBC_2021}.
To maintain the temperature of the system, we have used Dissipative Particle Dynamics (DPD) thermosat~\cite {Kremer_DPD}.
By writing the force $\mathbf{F}_i$ in an extended form, including terms from the DPD thermostat, Eq.~\ref{eq:sllod}b can be written as
\begin{equation}
    \dot{\mathbf{p}}_i = \sum_{j \neq i} \left( \mathbf{F}^c_{i,j} + \mathbf{F}^d_{i,j} + \mathbf{F}^r_{i,j} \right) - \dot{\gamma} \cdot p_{z,i} .
    \label{eq:dpd}
\end{equation}  
Here, $\mathbf{F}^c_{i,j} $ represents the conservative forces between particles $i$ and $j$. $\mathbf{F}^d_{i,j} = -\xi \omega(r_{ij}) (\mathbf{v}_{ij} \cdot \hat{\mathbf{r}}_{ij}) \hat{\mathbf{r}}_{ij} $ is the dissipative force, where $\xi$ is the friction coefficient, $\omega(r_{ij})$ is a weight function, and $\mathbf{v}_{ij}$ and $\hat{\mathbf{r}}_{ij}$ are the relative velocity and unit vector between particles $i$ and $j$, respectively. $ \mathbf{F}^r_{i,j} = \sigma \omega(r_{ij}) \theta_{ij} \hat{\mathbf{r}}_{ij} $ is the random force, where $\sigma$ controls the magnitude, $\theta_{ij}$ is a random number with zero mean, and $\omega(r_{ij})$ is the same weight function as used for $ \mathbf{F}^d_{i,j} $. 
To avoid disturbing or correcting velocities in the shear direction, and to make them comparable at different shear rates, the DPD thermostat is also altered to act only in the vorticity direction (namely $\xi=\left[0,\xi_y,0\right]$ and $\mathbf{F}^r=\left[0,f^r_{y},0\right]$)~\cite{LJ-Ruiz-Franco2018}.

\subsection{The Ewald Summation in Lees-Edwards Boundary Conditions}\label{subsec:the-ewald-summation-in-lees-edwards-boundary-conditions}
\begin{figure}[ht]
    \begin{center}
        \includegraphics[width=0.7\linewidth]{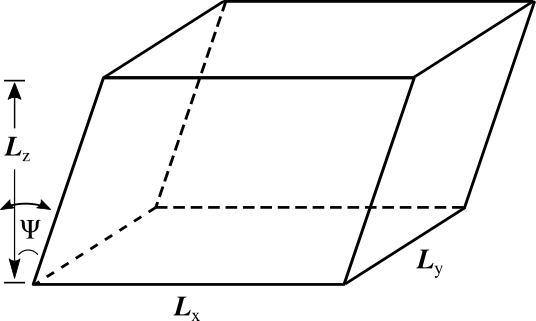}
        \caption{A parallelepiped box with a tilt angle $\psi$ at the shear direction.}
        \label{fig:parallelepiped}
    \end{center}
\end{figure}
In this work, all particles in the IL system are modeled using partial point charges and an infinite lattice of a unit cell is considered for determining the potential energy of long-range electrostatics interactions.
Then, the total potential energy of Coulomb interactions (in vacuum) can be written as 
\begin{equation}
    \begin{aligned}
        E_\mathrm{total}=\frac{1}{4\pi\epsilon_0}\sum_{i,j}\frac{q_iq_j}{|r_{i,j}|+n\mathbf{L}}
    \end{aligned}
    \label{eq:ecoul}
\end{equation}
where $\epsilon_0$ is the vacuum permittivity, $q_i$ and $q_j$ are the partial charges of particles $i$ and $j$ with their distance $r_{i,j}=|\mathbf{r}_i-\mathbf{r}_j|$ and $n$ is an arbitrary integer triplet which forms a repeat vector of $n\mathbf{L}$.
For such a model system, the Ewald summation can be split into real- and reciprocal-space contributions with a correction term:
\begin{equation}
    \begin{aligned}
        U_\mathrm{Ewald}=U_\mathrm{real}+U_\mathrm{recip.}+U_\mathrm{corr.}.
    \end{aligned}
    \label{eq:ewald}
\end{equation}
Based on an orthogonal unit cell, the real-space part
\begin{equation}
    \begin{aligned}
        U_\mathrm{real}=\sum_{i\neq j}\frac{q_iq_j}{4\pi\epsilon_0}\frac{\mathrm{erfc}(\alpha|r_{i,j}|)}{|r_{i,j}|}
    \end{aligned}
    \label{eq:rspace}
\end{equation}
is considered as short-range interactions under a spherical cutoff of $r_\mathrm{cut}$. Here, $\alpha$ is the converging parameter and erfc() refers to the complementary error function.
In \epp, $U_\mathrm{real}$ is similarly computed with LEbc to that of other short-range interactions (e.g. Lennard-Jones, bonded interactions etc.) as described in Ref.~\cite{ESPP_LEBC_2021}.

Focusing on the reciprocal part, we consider the simplified expression for $U_\mathrm{recip.}$ under the assumption of point charges in orthogonal unit cell images, which allows the form:
\begin{equation}
    \begin{aligned}
        U_\mathrm{recip.}=\frac{1}{2V\epsilon_0}\sum_{\mathbf{k}\neq 0}^{\mathbf{k}_\mathrm{max}}\frac{e^{-k^2/(4\alpha^2)}}{k^2}|\sum_{j}q_je^{i\mathbf{k}\cdot\mathbf{r}_j}|^2
    \end{aligned}
    \label{eq:kspace}
\end{equation}
Here, $\mathbf{k}$ is a reciprocal $k$-vector obtained from the Fourier transform of the charge density field:
\begin{equation}
    \begin{aligned}
        \mathbf{k}=\begin{bmatrix}
                    2\pi k_x/L_x \\
                    2\pi k_y/L_y \\
                    2\pi k_z/L_z
        \end{bmatrix}
    \end{aligned}
    \label{eq:kvec}
\end{equation}
where $k_x\in[0,k_\mathrm{max}]$ and $k_y$ and $k_z\in[-k_\mathrm{max},k_\mathrm{max}]$ but excluding the zero vector of $\mathbf{k}=\left[0,0,0\right]$.

When applying a shear flow using LEbc, the conventional method of constructing the periodic lattice by duplicating unit cell images must be adapted.
For that, Wheeler \emph{et al.} presented a generalized form of Ewald summation for the unit cell with an arbitrary parallelepiped~\cite{10.1080/002689797170608}.
Figure~\ref{fig:parallelepiped} presents a simplified parallelepiped by only tilting the $xz$-plane in the shear direction and the tilt angle $\psi$ always develops between $-\tan^{-1}(L_x\slash L_z)$ and $+\tan^{-1}(L_x\slash L_z)$ during the simulation.
The total coulomb potential (Eq.~\ref{eq:ecoul}) is then modified by introducing a cell basis matrix $\bfit{D}$:
\begin{equation}
    \begin{aligned}
        E_\mathrm{total}=\frac{1}{4\pi\epsilon}\sum_{i,j}\frac{q_iq_j}{|r_{i,j}|+\bfit{D}\cdot{n\mathbf{L}}}.
    \end{aligned}
    \label{eq:ecoul-D}
\end{equation}
In practice, $(\bfit{D}^T)^{-1}$ is computed with the given $\psi$:
\begin{equation}
    \begin{aligned}
    (\bfit{D}^T)
        ^{-1}=\begin{pmatrix}
                  1/L_x         & 0     & 0     \\
                  0             & 1/L_y & 0     \\
                  -\tan(\psi)/L_x & 0     & 1/L_z
        \end{pmatrix} ,
    \end{aligned}
    \label{eq:dmatrix}
\end{equation}
and a new $k$-vector is generated by multiplying the original one with $(\bfit{D}^T)^{-1}$:
\begin{equation}
    \begin{aligned}
        \mathbf{k}_\psi&=(\bfit{D}^T)^{-1}(\mathbf{k}\cdot \mathbf{L})\\
        &=\begin{bmatrix}
              2\pi k_x/L_x \\
              2\pi k_y/L_y \\
              2\pi \left(\dfrac{k_z}{L_z}-\tan(\psi)\dfrac{k_x}{L_x}\right)
        \end{bmatrix} .
    \end{aligned}
    \label{eq:kvec-D}
\end{equation}
It is used to replace Eq.~\ref{eq:kvec}.

The current studied system consists of multi-site molecules and has a neutral net charge.
Hence, the correction term, $U_\mathrm{corr}$, includes both self-interaction exclusion and a coulomb correction for intra-molecular interactions (for which an exclusion rule applies):
\begin{equation}
    \begin{aligned}
        U_\mathrm{corr}=-\sum_{j}^\mathrm{self}\frac{\alpha}{4\pi\epsilon_0\sqrt{\pi}}q_j^2-\sum_{i\neq j}^\mathrm{excl.}\frac{q_iq_j}{4\pi\epsilon_0}\frac{\mathrm{erf}(\alpha|r_{i,j}|)}{|r_{i,j}|}.
    \end{aligned}
    \label{eq:ecorr}
\end{equation}
Combining Eq.~\ref{eq:rspace}, \ref{eq:kspace} and \ref{eq:ecorr} into Eq.~\ref{eq:ewald} gives the exact Ewald sum for the IL system.
We also need to note that it is possible to calculate the real-space contribution to the Ewald sum in a periodic system with triclinic deformation, following the reference work~\cite{10.1080/002689797170608} by Wheeler \emph{et al.}
In \epp, however, the orthogonality of a simulation box is retained with the LEbc implementation.
Instead of deforming the unit cell itself, only certain image cells---specifically, those adjacent in the shear direction---are sheared according to the imposed velocity gradient.
The simulation box and image cells still form an infinite lattice, but with staggered vertical joints.
\epp~already includes a force calculator capable of handling short-range interactions within this specialized staggered geometry. As a result, there is no immediate need to compute the real-space Ewald term using a deformation matrix $\bfit{D}$, which would otherwise require additional implementation effort.

Note that when implementing LEbc in MD simulations with domain decomposition, it is necessary to ensure that the box length in the shearing direction (here, the $x$-direction) is larger than $\max(r_c + r_b,~ 2r_c)$, where $r_c$ is the cut-off distance and $r_b$ is a small buffer added beyond the cut-off to allow for a smooth decay of the interaction potential. This is in addition to the usual requirement that the box length be greater than $r_c + r_b$. This condition ensures that no interactions are missed when adjacent cells are shifted due to the shearing of neighboring boxes.

\subsection{Shear Viscosity}\label{subsec:shear-viscosity}

In many applications of ILs, such as lubrication, it is essential to understand their viscous behavior and how their viscosity changes during processing. The shear viscosity $ \eta $ can be computed at a finite shear rate as
\begin{equation}
    \begin{aligned}
        \eta=\frac{\left<\varsigma_{xy}\right>}{\dot{\gamma}} , 
    \end{aligned}
    \label{eq:viscosity}
\end{equation}
where $ \varsigma_{xy} $ denotes the off-diagonal component of the shear stress tensor (\bfit{$\varsigma$}).
The shear stress tensor subject to LEbc can be calculated from the Irving-Kirkwood model~\cite{Irving1950}:

\begin{equation}
    \begin{aligned}
        \varsigma=-\dfrac{1}{V}\left[\sum_{i} \left(\mathbf{p}_i\otimes\mathbf{p}_i/m_i\right)+\sum_{i}\sum_{j (j>i)}\mathbf{r}_{i,j}\otimes \mathbf{F}^c_{i,j}\right],
    \end{aligned}
    \label{eq:tensor}
\end{equation}
where $ \otimes $ denotes the dyadic product.
More generally, $N$-body interactions ($N\ge3$) can be first decomposed into a sum of pair interactions and the corresponding stress tensors can be thus calculated using Eq.~\ref{eq:tensor}.
For electrostatic interactions treated with the EW method, the short-range component corresponds to the the pairwise term $\mathbf{F}^c_{i,j}$ in Eq.~\ref{eq:tensor}. The long-range contributions to the viscosity can be determined following the approach implemented in DL\_MESO~\cite{DL-MESO-Seaton}, which involves computing the long-range contribution to the virial in a deformed periodic space.


\section{Simulation Details}\label{sec:simulation-details}
\begin{figure}[ht]
    \begin{center}
        \includegraphics[width=0.8\linewidth]{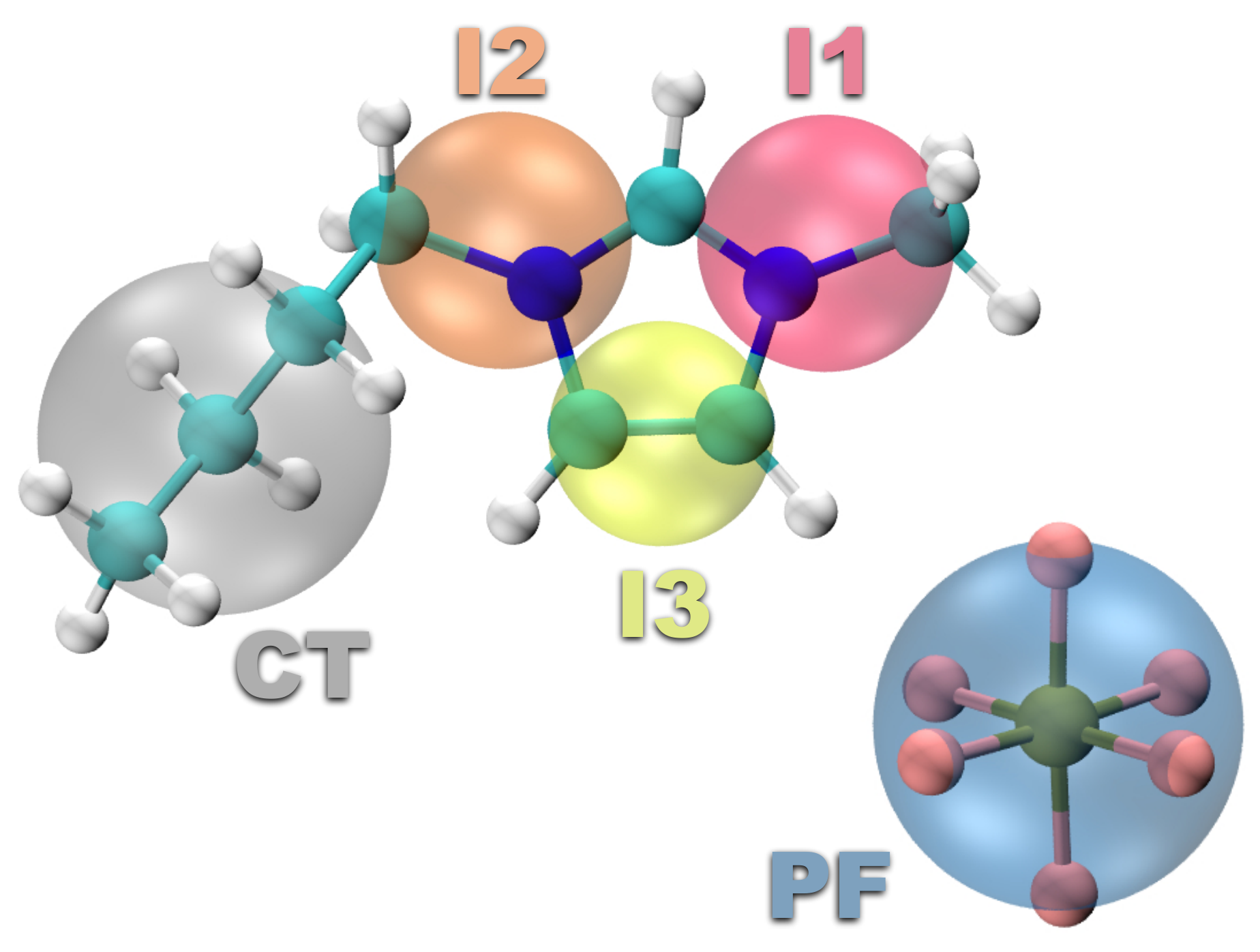}
        \caption{(Republished from~\cite{Rudzinski_2021}) In CG representation, each ion is partitioned into some atomic groups, as indicated by the large transparent spheres: 4 for the cation and 1 for the anion. All CG beads were mapped to the center of mass of the corresponding group of atoms: (i) CT—the last 3 carbon groups of the alkyl chain; (ii) I2—the first carbon group of the alkyl chain, the 1-nitrogen of the ring, and ‘half’ of the 2-carbon group of the ring; (iii) I1—the 3-nitrogen and the associated methyl group of the ring and also ‘half’ of the 2-carbon group of the ring; (iv) I3—the 4- and 5-carbon groups of the ring; (v) PF—the entire anion.}
        \label{fig:mapping}
    \end{center}
\end{figure}

This work considers the imidazolium-based IL [C$_4$MIM]$^+$[PF$_6$]$^-$, i.e., a 1-butyl-3-methylimidazolium cation (with the chemical formula $\text{C}_8\text{H}_{15}\text{N}_2^+$) and a hexafluorophosphate anion (with the chemical formula $\text{PF}_6^-$).
Following our previous work~\cite{Rudzinski_2021}, we employ a CG model that represents each cation with 4 CG sites and each anion with a single
CG site. The imidazolium ring is represented by 3 sites, I1,
I2 and I3, mapped to the center of mass of a corresponding group of atoms, as illustrated with the large transparent
spheres in Fig.~\ref{fig:mapping}. Note that the I1 and I2 sites overlap, sharing
contributions from the 2-carbon of the five-membered ring
(i.e., the carbon flanked by two nitrogens). The butyl chain
is represented by an additional site, denoted CT, while the
anion site is denoted PF.

The interactions for the CG model were also adapted from our previous work~\cite{Rudzinski_2021}. In particular, the ``struct'' model, which was parametrized to reproduce the individual inter- and intra-molecular distributions along the order parameters describing each CG interaction, was employed with a pressure correction added to the intermolecular CG potentials. This correction was relatively small and resulted in near-negligible changes in the corresponding radial distribution functions while enabling NPT simulations with $\sim 0.2 \%$ deviation in density relative to the reference atomistic model. The tabular force field files are openly available as described in Section~\ref{sec:data-manag} and in the Data Availability section.

All simulations considered in this work contain 550 ion pairs in a rectangular box of size $6.123\text{nm} \times 5.525\text{nm} \times 5.525\text{nm}$, satisfying the domain decomposition rules with the given cutoff.
This results in a density identical to the results with the atomistic force field from Bhargava and Balasubramanian~\cite{bhargava_refined_2007}, from which the CG model was derived.
Unless otherwise specified, all simulations presented were performed at a temperature of 300~K.

Following previous work~\cite{Rudzinski_2021}, the original all-atom (AA) configuration was obtained using a standard equilibration protocol: an NPT simulation employing the Nosé–Hoover thermostat~\cite{10.1063/1.447334, PhysRevA.31.1695} and the Parrinello–Rahman barostat~\cite{10.1063/1.443248} to determine the box size and system density at the target temperature, followed by several NVT simulations to relax the system, all performed using the Gromacs software package. 
Subsequently, an equilibrated configuration was mapped to the CG representation and then energy-minimized according to the CG force field.
From the initial configurations, a 10~ns NVT equilibration simulation was performed at equilibrium (i.e., $\dot{\gamma} =$ 0~ns$^{-1}$), using the Gromacs simulation package for efficiency purposes. 
From these simulations, we then selected three configurations from the end of the trajectory, each 1~ns apart, as starting configurations for the production shear flow runs with \epp{}. During this time interval, an ion moves approximately 0.5–0.6 nm, according to the diffusion coefficient calculated for the system, as shown later in Fig.~\ref{fig:msd}.

NVT production simulations were performed with both EW (with $k_\mathrm{cutoff} = 15$) and RF electrostatics at various shear rates (ranging from 0.01\,ns$^{-1}$ to 10\,ns$^{-1}$).
Each of the production simulations was run for 10\,ns with a time step of 1\,fs and the trajectory was written out every 0.1\,ps.
The DPD thermostat was used to couple the velocities in the vorticity direction, i.e.,\ the $y$-direction for the present systems, with a friction coefficient of $1\,$ps$^{-1}$. The cutoff for coulomb interactions (real-space) and Lennard-Jones interactions ($r_c$) was set to 1.5\,nm, with a small buffer distance ($r_b$) of 0.3 \,nm to allow for a smooth decay to zero in the calculation of interaction potentials.
Additionally, with the same parameters, equilibrium simulations were performed at 300, 320, and 340\,K to later compare the results.

\section{Data Management}\label{sec:data-manag}

Storage and transparency of the simulation data generated in this study is achieved using the NOMAD software and repository~\cite{scheidgen2023nomad}. NOMAD [nomad-lab.eu] is an open-source, community-driven data infrastructure, built following the FAIR principles~\cite{Wilkinson2016FAIR}. NOMAD extracts extensive metadata from raw simulation files and stores them within a structured and \emph{normalized} (i.e., code-agnostic) schema. As a result, all details of the simulations can be easily accessed and browsed. In addition to support for some of the largest molecular dynamics packages (e.g., Gromacs and Lammps), NOMAD enables the upload of any molecular dynamics data via an specialization of the H5MD schema, called H5MD-NOMAD (see the NOMAD documentation for more details~\cite{NOMADDocs}).

The ESPResSo++ simulation software functions through execution of a python script that records simulation trajectories and energy outputs according to a user-defined format (Gromacs \textit{.xtc} and \textit{.dat} format in this study).
To prepare the data for upload to NOMAD, a script for mapping from the raw ESPResSo++ output to the H5MD-NOMAD format was prepared. All input parameters were manually curated into json files, which were then directly mapped to the H5MD file. 

For each simulation, the raw input files (ESPResSo++ execution script and configuration file), output files (including energy outputs, the results \textit{.h5} file containing the parameters presented in this work, the trajectory file, the velocity file, and trimmed log files from the simulation execution on HPC resources), and the generated H5MD-NOMAD file were compressed to a zip file and then uploaded to NOMAD and published using the `nomad-utility-workflows` module~\cite{Rudzinski2025Utility}. The tabular force field files and Gromacs-style topology file (used for definition of charges and molecular connections), identical for all simulations, were separately uploaded for storage efficiency, and linked to each entry through the upload metadata. The upload metadata was also edited to include additional overarching information, e.g., molecular characterization of the system. Finally, a dataset was created to group the simulations and retrieve a DOI: \url{https://dx.doi.org/10.17172/NOMAD/2025.04.15-1}

\section{Results}\label{sec:results}

In this section, we present a detailed analysis of the behavior of a CG model of the IL [C$_4$mim]$^+$ [PF$_6$]$^-$ under shear, explicitly assessing the impact of electrostatics by comparing simulations using EW and RF treatments. We begin with a validation of the simulation framework and parameters. This is followed by an examination of the system's structural organization, focusing on pairwise structure and molecular orientation. We then investigate dynamical properties at varying time/length scales via incoherent intermediate scattering functions (short-intermediate/molecular), as well as mean squared displacements and diffusion constants (intermediate-long/molecular). Finally, we evaluate the dynamic heterogeneity of the system using the non-Gaussian parameter. For each of these properties, we determine a critical shear rate, before which they remain relatively insensitive, and beyond which they exhibit significant changes.

\subsection{Validation}\label{subsec:validation}
\begin{figure}[ht]
    \begin{center}
        \includegraphics[width=0.9\linewidth]{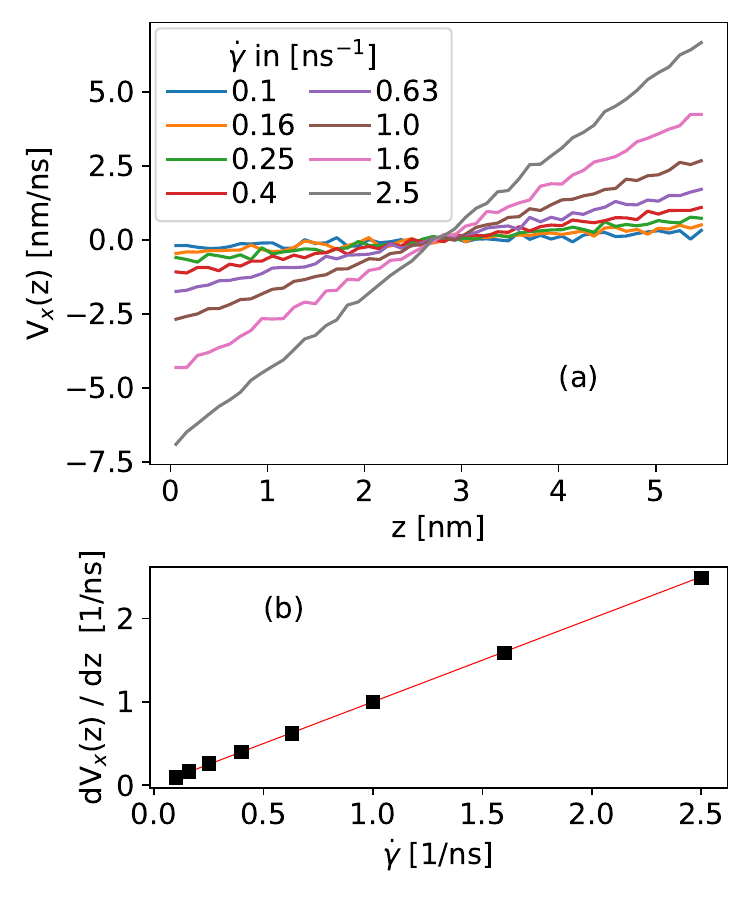}
        \caption{(a) Velocity profiles for different shear rates obtained using the EW method. A similar profile is observed in the RF simulations. (b) Slope of the linear fits to the velocity profiles as a function of shear rate. The red line corresponds to the identity function.}
        \label{fig:v_x_over_z_ew}
    \end{center}
\end{figure}

As described in Section~\ref{sec:methods}, the shear force applied during the simulations should generate a profile of the velocity in the $x$ direction as a function of the $z$ coordinate that is linear on average with $v_x(z=0) = 0$ for simple homogeneous liquid. Fig.~\ref{fig:v_x_over_z_ew}(a) demonstrates this trend, with a systematically increasing profile slope as a function of increasing shear rate, as expected. Panel (b) of Fig.~\ref{fig:v_x_over_z_ew} shows the slope of the linear fits to the velocity profiles, which correspond to the shear rates applied to the system.

\begin{figure}[ht]
    \begin{center}
        \includegraphics[width=0.9\linewidth]{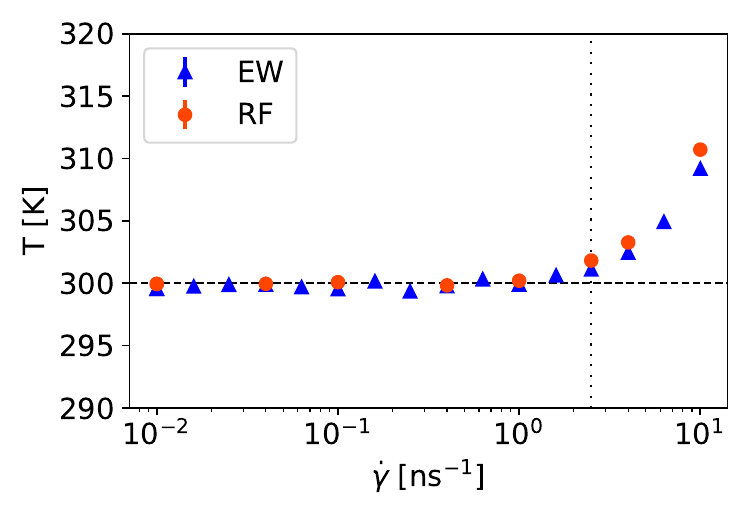}
        \caption{The mean temperature for the last nanoseconds of each simulation over the shear rate compares the EW method (blue triangles) with the reaction field (RF, red circles). The values shown are averaged over three simulations for each shear rate. The dashed horizontal line indicates the target temperature of \(300\,\text{K}\), and the dotted vertical line shows the upper limit for the range of shear rates considered in the remainder of this work.}
        \label{fig:mean_T_all_rates}
    \end{center}
\end{figure}

The velocity added to the $x$-direction will dissipate throughout the system and, ideally, the corresponding extra energy will be transferred out of the system via the thermostat. However, in practice, given a fixed coupling constant, the thermostat will only be able to effectively extract this surplus of energy until a shear rate threshold. 
Beyond this shear rate, the applied shear forces will overpower the thermostat, resulting in an increase of the average simulation temperature.
Fig.~\ref{fig:mean_T_all_rates} demonstrates this concept by presenting the mean temperature of the last nanosecond of the production simulation, further averaged over 3 independent simulations, for each of the simulated shear rates.
To avoid unintended system heating, the remainder of this work considers shear rates up to $\dot\gamma = 2.5\,$ns$^{-1}$ (vertical dotted line in Fig.~\ref{fig:mean_T_all_rates}). The shear rate and treatment of electrostatics are varied systematically in distinct simulations for analysis.
All other simulation control parameters, including the thermostat coupling constant remain identical to ensure consistent comparisons throughout.
We note that for the highest shear rates considered, the RF simulations produce slightly higher temperatures than corresponding EW simulations. However, we do not expect this to affect the conclusions reached from these comparisons.

\begin{figure}[ht]
    \begin{center}
        \includegraphics[width=0.9\linewidth]{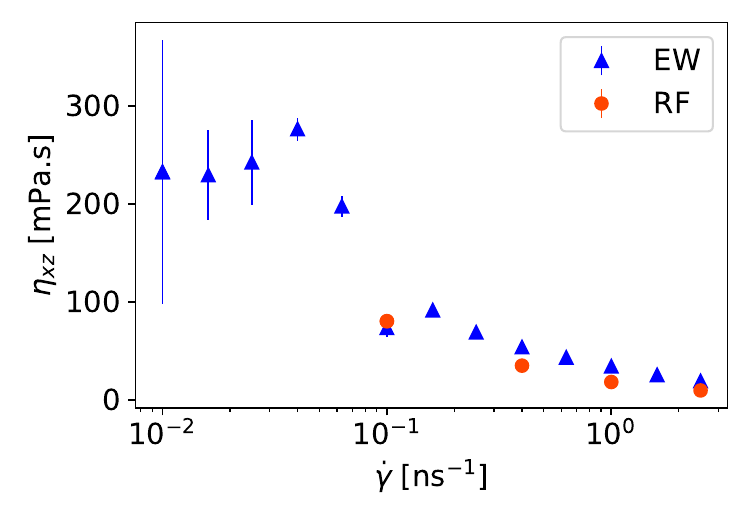}
        \caption{Shear viscosity over shear rate for simulations with EW (blue) and RF (red). Values are calculated from three different runs.}
        \label{fig:shear_vis}
    \end{center}
\end{figure}

To conclude the validation, we examine the transport-level collective behavior via the shear viscosity. Fig.~\ref{fig:shear_vis} presents the shear viscosity in the shear plane, \(\eta_{xz}\), calculated according to Eqs.~\ref{eq:viscosity} and \ref{eq:tensor}, as a function of shear rate for both EW (blue markers) and RF (red markers) simulations.
In the case of EW, the viscosity remains relatively insensitive at low shear rates before undergoing a systematic change beyond a critical shear rate, \(\dot\gamma \sim 4 \times 10^{-2}\)~ns\(^{-1}\). This result is a demonstration of the well-known ``shear thinning'' effect. However, the off-diagonal elements of the stress tensor are quite noisy, resulting in a rather large uncertainty of the viscosity values, especially at lower shear rates. We present the viscosity for the RF simulations for completeness, but leave out the low-shear values calculated as we found them to be completely unreliable. This is due to the slower dynamics of ions in the RF simulations (discussed further below). In the following, we determine the critical shear rate more accurately, and for both electrostatic approaches, via other structural and dynamical properties. 

\subsection{Structure}\label{subsec:structural-changes}
The structure of the liquid is first assessed via the pairwise radial distribution function (RDF):
\begin{equation}
    g(r)=\frac{1}{n_j \cdot N_i}\sum_i\sum_j\frac{\delta(r_{ij}-r)}{4\pi r^2}, \label{eq:rdf}
\end{equation}
with the distance $r_{ij} = |\vec r_i-\vec r_j|$ between particle $i$ and $j$, the number density $n_j$ of particle $j$, and the total number $N_i$ of particles $i$ over which is summed.
 
\begin{figure}[ht]
    \centering
    \includegraphics[width=0.9\linewidth]{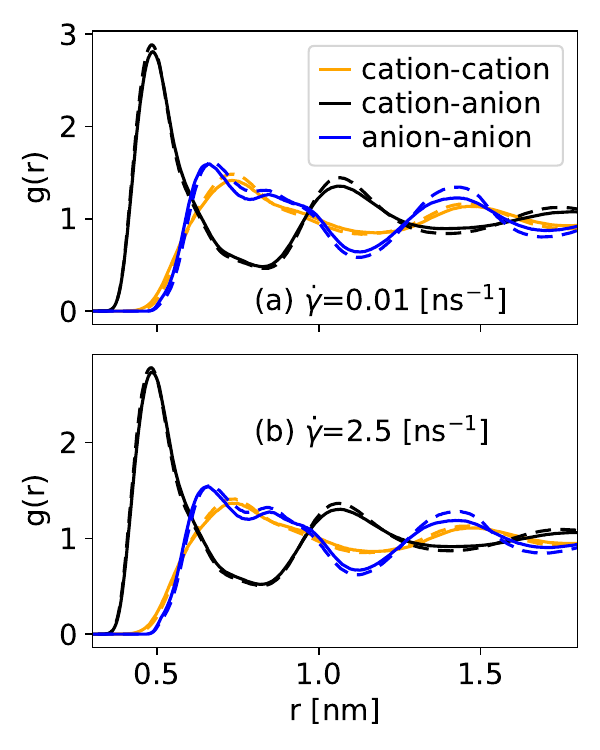}
    \caption{Comparison of the different RDFs at a) low shear rate (0.01\,ns$^{-1}$) and b) high shear rate (2.5\,ns$^{-1}$). Solid lines are for simulations with EW method, and dashed lines are for the RF. Error bars (not shown) are smaller than the line's thickness.}
    \label{fig:rdf_systems}
\end{figure}

Fig.~\ref{fig:rdf_systems} presents cation-cation (yellow curves), cation-anion (black curves), and anion-anion (blue curves) molecular center-of-mass RDFs for simulations performed with EW (solid curves) and RF (dashed curves) electrostatics. Panels (a) and (b) present results from simulations with shear rates of $\dot{\gamma} = $0.01\,ns$^{-1}$ and $\dot{\gamma} = $ 2.5\,ns$^{-1}$, respectively.
At the lower shear rate (panel a), there are visible differences between the two electrostatic treatments, especially within the 2nd and 3rd solvation shell peaks. In particular, the RF RDFs (dashed curves) exhibit enhanced local structural order compared with EW. 
The averaging over long-range electrostatics within the RF approach apparently shifts the balance between the interactions stabilizing local structures.
This result indicates the potential importance of accurate long-range electrostatic treatments for this charged system. The magnitude of these discrepancies are clearly diminished at the higher shear rate (panel b). 

\begin{figure}[ht]
    \centering
    \includegraphics[width=0.9\linewidth]{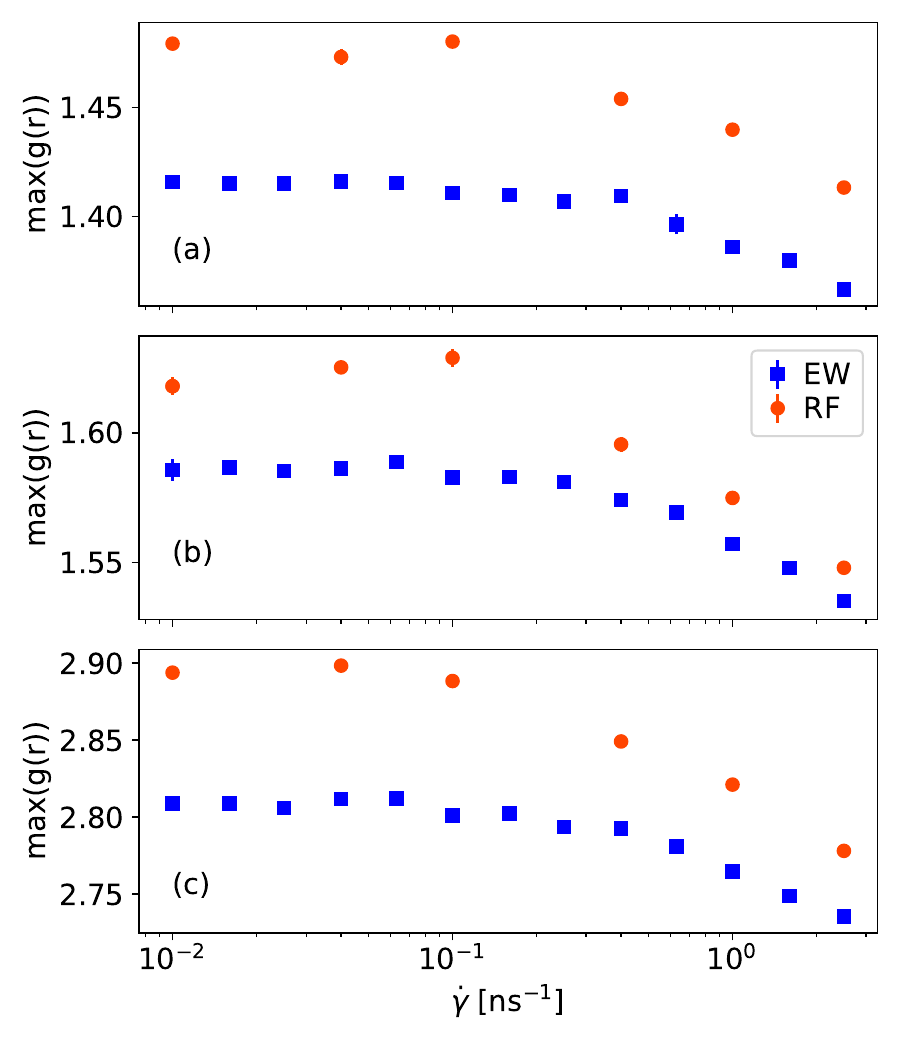}
    \caption{The height of the nearest-neighbor RDF peak as a function of shear rate for a) cations, b) anions and c) cation-anion pairs with EW (blue markers) and RF (red markers) electrostatics.}
    \label{fig:rdf_peak}
\end{figure}

Figs.~S1 and S2 in the supporting material present a comparison of the same 3 RDFs over the range of examined shear rates, for EW and RF electrostatics, respectively. However, because the RDF changes very subtely as a function of shear, it is difficult to determine systematic behavior from these plots.
Instead, to quantitatively assess these changes, panels (a), (b), and (c) of Fig.~\ref{fig:rdf_peak} present the height of the first peak of the cation-cation, cation-anion, and anion-anion RDFs, respectively, as a function of shear rate for both EW (blue markers) and RF (red markers) simulations. All three RDF peak heights remain relatively constant until a critical shear rate, $\dot\gamma \sim 10^{-1}\,$ns$^{-1}$, before decreasing systematically and approximately linearly. Thus, the shear thinning observed for the shear viscosity (Fig.~\ref{fig:shear_vis}) is accompanied by an enhanced disorder in the local IL structure.

\begin{figure}[ht]
    \centering
    \includegraphics[width=\linewidth]{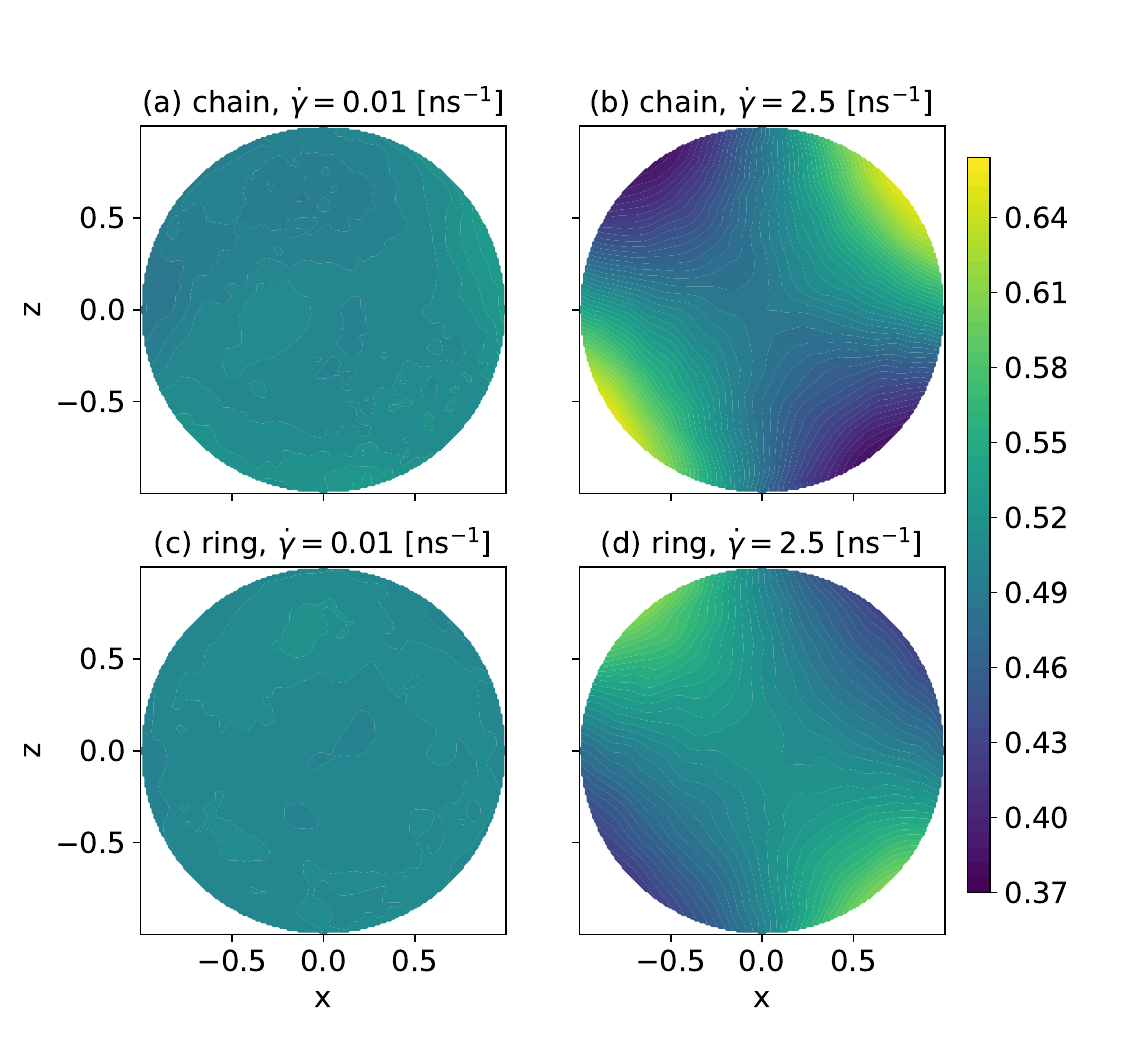}
    \caption{Orientation probability for the I2-CT vector (a and b) and ring-normal vector (c and d) for low shear rate 0.01\,ns$^{-1}$ (a and c) and high shear rate 2.5\,ns$^{-1}$ (b and d). Data is averaged from 3 simulations with the EW method. The probability data are normalized in a way that they integrate to one.}
    \label{fig:orientation}
\end{figure}

Higher-order structural arrangements were assessed by investigating the orientational preferences of the cation.
In particular, we considered 2 orientational vectors: 1. the alkyl chain end-to-end vector represented by I2 and CT beads ($\overrightarrow{\text{I2~CT}}$, denoted ``chain''), and 2. the normal vector of the imidazole ring represented by the I1, I2 and I3 beads ($\overrightarrow{\text{I1~I2}} \times \overrightarrow{\text{I1~I3}}$, denoted ``ring'').
Fig.~\ref{fig:orientation} presents the probability density of the $x$ and $z$ components of these orientational vectors.
The top (a / b) and bottom (c / d) panels show the results for the chain and ring vectors, respectively. The left (a/c) and right (b/d) panels present results for shear rates of $\gamma = $ 0.01\,ns$^{-1}$ and 2.5\,ns$^{-1}$, respectively.

Panels (a) and (c) demonstrate no preferred orientation for low shear rates, as expected based on previous investigations of this IL~\cite{Rudzinski_2021, Vogel_JCP_19_TP, kloth_coarse-grained_2021}. In contrast, at the higher shear rate (panels b and d), there is a strong orientational preference for both vectors in the $xz$-plane. As suggested by the geometry of the cation, these preferred orientations are nearly orthogonal, with the chain and ring vectors forming angles of approximately \( 36^\circ \) and \( 51^\circ \) with the $x$-axis when projected onto the $xz$-plane. The magnitude of the orientational probabilities also indicates that the alkyl chain (panel b) is more aligned than the ring-normal vector (panel d). This is likely due to the flexibility of the chain, in contrast to the ring's more rigid structure with greater influence from steric and electrostatic interactions.

\begin{figure}[ht]
    \centering
    \includegraphics[width=0.9\linewidth]{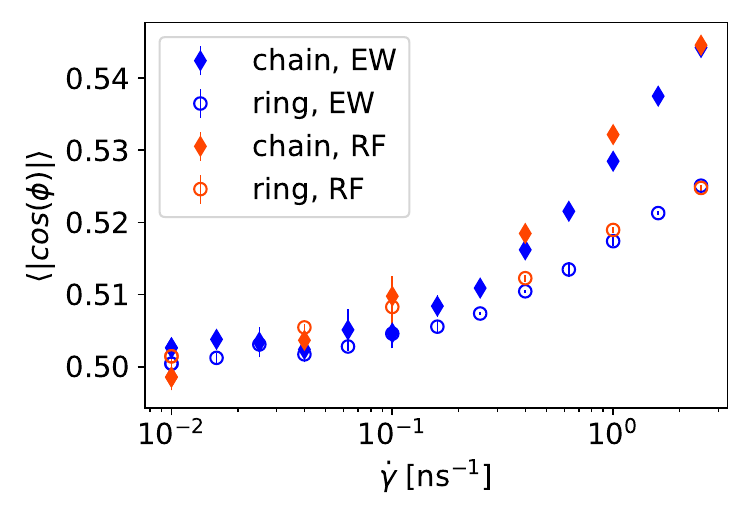}
    \caption{Average absolute cosine of the angle between each orientational vector and its distribution peak. Chain and ring vectors are shown with diamond and circle markers, with EW and RF results in blue and red, respectively.}
    \label{fig:orientation_to_vector}
\end{figure}

To further quantify these orientational preferences, we considered the angle formed between each orientational vector and the vector pointing to the position of maximum probability of the distribution shown in Fig.~\ref{fig:orientation}(b) and (d) (preferred orientation at high shear rate). 
Fig.~\ref{fig:orientation_to_vector} presents the average absolute value of the cosine of this angle 
as a function of shear rate, for both chain (diamond markers) and ring (circle markers) orientational vectors as well as for both EW (blue markers) and RF (red markers) electrostatics. 
At equilibrium, the orientational vectors are nearly uniformly distributed, resulting in \(|\cos(\phi)|\) values evenly spread between 0 and 1, with an average of approximately 0.5. However, when shear forces are applied, the vectors tend to align more with the preferred orientation, leading to a non-uniform distribution of \(|\cos(\phi)|\) that skews toward 1. Consequently, the average value of this function increases toward 1, representing the scenario in which all vectors are perfectly aligned. Overall, the trends are similar across all cases: small or negligible changes until a critical shear rate of $\dot\gamma \sim 10^{-1}\,$ns$^{-1}$, 
followed by a consistent increase in orientational alignment. This (inverse) timescale is clearly shorter than the correlation times of these vectors, which are approximately 1 ns and 0.08 ns for the chain and ring vectors at equilibrium, respectively (not shown). Moreover, the stronger alignment of the alkyl-chain vector upon the onset of shear, already mentioned above, is more clearly visible from these results. The critical shear rate determined from this orientational analysis aligns perfectly with the changes observed in the first peak of the RDF, indicating that a disruption of the intrinsic local structure of the IL accompanies alignment with the flow field.

\subsection{Dynamics}\label{subsec:dynamical-changes}

We first assessed the dynamics at short-to-intermediate timescales and on short (nearest-neighbor) length scales using the incoherent intermediate scattering function (ISF), generally written as
\begin{equation}
    S_{\vec{q}}(t ) = \left\langle \frac{1}{N} \sum\limits_{i=1}^N \exp[-\mathrm{i}\vec{q}(\vec{r}_i(t_0 + t )-\vec{r}_i(t_0))]\right\rangle_{t_0}, \label{eq:isf}
\end{equation}
where $q=11.8$\,nm$^{-1}$ is the modulus of the scattering vector, $N$ is the number of ions, and the angle brackets indicate the average over various time origins, $t_0$.
In the following, we only consider motion along the $y$-axis, i.e., in the direction orthogonal to the shear. This provides a clearer result by effectively disregarding trivial displacements that arise from shearing. Thus, the ISF can be simplified to:
\begin{equation}
    S_{q, y}(t) =  \left\langle \frac{1}{N} \sum\limits_{i=1}^{N} \cos(q|y(t_0+t) - y(t_0)|)  \right\rangle_{t_0}. \label{eq:isf_1d}
\end{equation}
Moreover, we calculate the ISF as a function of the center of mass position of each ion.
Each calculated ISF curve was fitted with a Kohlrausch-Williams-Watts (KWW) function to extract correlation times $\tau$.
The KWW function is defined as:
\begin{equation}
    S(t) = A \cdot \exp \left[-\left(\frac{t}{\tau}\right)^\beta\right] , \label{eq:kww}
\end{equation}
where $A$ is the amplitude parameter, $\tau$ is the correlation time, and $\beta$ is the stretching parameter.

\begin{figure}[ht]
    \begin{center}
        \includegraphics[width=0.9\linewidth]{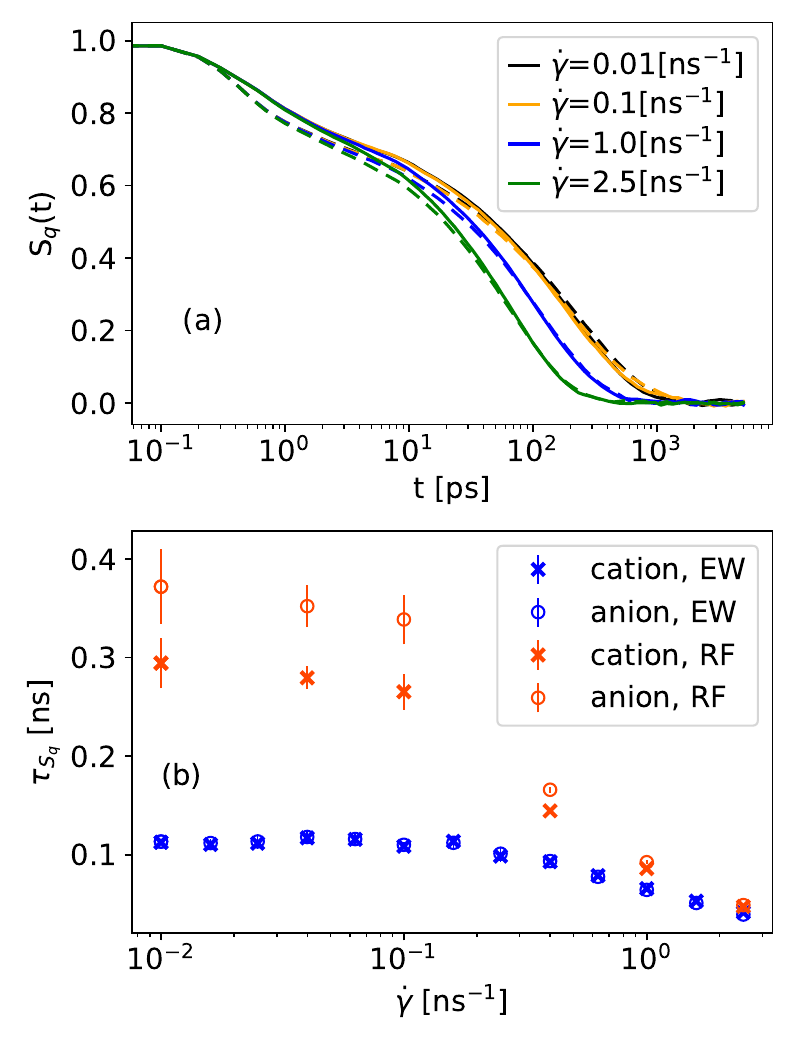}
        \caption{a) 
        Decay of the incoherent intermediate scattering function (ISF) of the cation (solid lines) and anion (dashed lines) compared at different shear rates for the simulations with EW electrostatics. b) Shear rate dependence of the correlation times $\tau$ from the ISF of the cation and anion for both simulations with the EW method and RF.}
        \label{fig:isf}
    \end{center}
\end{figure}

Panel (a) of Fig.~\ref{fig:isf} presents the ISF curves for cations (solid lines) and anions (dashed lines) across a range of shear rates for the EW simulations only. The ISF curves for simulations with RF electrostatics are presented in the supporting material in Fig.~S3. The ISF exhibits two distinct features: an initial vibrational decay at short timescales, followed by a slower relaxation process referred to as the main process. The vibrational decay is associated with high-frequency molecular vibrations and remains unaffected by the shear. In contrast, the main process, which corresponds to the structural relaxation of the system, shifts to shorter timescales as the shear rate increases. This trend indicates that higher shear rates accelerate the decorrelation of molecular positions in agreement with a decrease of viscosity as a function of increasing shear (Fig.~\ref{fig:shear_vis}).

Panel (b) of Fig.~\ref{fig:isf} presents the correlation times determined from the fitted ISF curves for both EW (blue markers) and RF (red markers) simulations as a function of the shear rate.
The EW simulations display approximately constant correlation times over lower shear rates, followed by a systematic decrease starting at a critical shear rate of $\dot\gamma \sim 10^{-1} \,$ns$^{-1}$. This result is quite consistent with the impact of shear on both the pairwise structure (Fig.~\ref{fig:rdf_peak}) and the orientational preference of the cation (Fig.~\ref{fig:orientation_to_vector}). 
The correlation times from the RF simulations also decrease as a function of increasing shear. However, the calculated times are quantitatively inconsistent with the EW results, with generally more sensitivity towards the application of shear even in the low-shear regime. 
The increased local order observed for the RF simulations in Fig.~\ref{fig:rdf_systems} is accompanied by a (more apparent) increase in ionic correlations times, i.e., a slowing of the dynamics. This dramatic difference in structural relaxation timescales indicates a fundamental limitation of the RF approach for IL simulations, even at the CG level of resolution. 

We characterize translational dynamics at intermediate to longer time and length scales by examining the cation and anion center-of-mass mean-square displacements (MSDs).
As with the ISF, only the displacements in $y$-direction are taken into account: 

\begin{equation}
    \langle \Delta y^2 \rangle(t) = \left\langle \frac{1}{N} \sum\limits_{i=1}^N \left[y_i(t_0 + t)-y_i(t_0)\right]^2 \right\rangle_{t_0},
    \label{eq:msd}
\end{equation}
where $y_i(t_0)$ and $y_i(t_0 + t)$ are the $y$-coordinate of the ion's center of mass at the initial and final times of a time interval $t$, $N$ is the total number of ions, and  $\langle \rangle_{t_0}$ represents average over various time origins $t_0$.

\begin{figure}[ht]
    \begin{center}
        \includegraphics[width=0.9\linewidth]{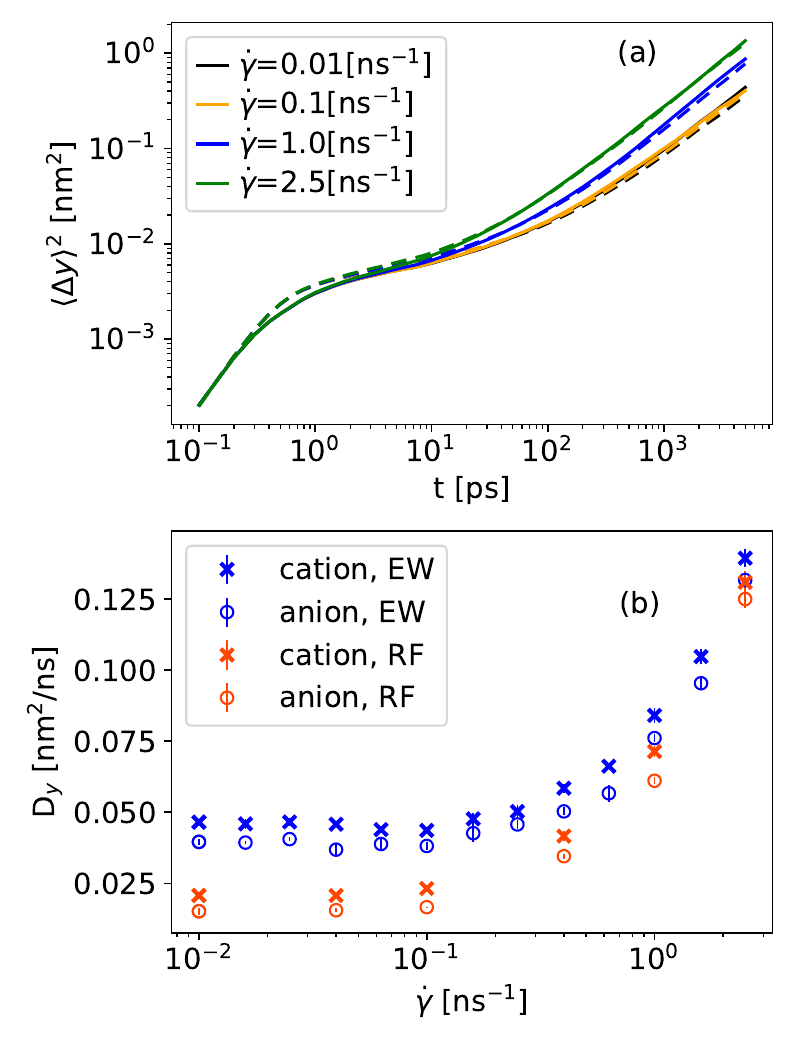}
        \caption{a) Mean square displacement (MSD) in the $y$-direction for the cations (solid lines) and anions (dashed lines) compared at different shear rates for the simulations using EW method. b) Shear rate dependence of the self-diffusion coefficient in $y$-direction compared between cation and anion as well as the two electrostatic treatments to calculate coulomb interactions.}
        \label{fig:msd}
    \end{center}
\end{figure}

Panel (a) of Fig.~\ref{fig:msd} presents the MSD curves for both cations (solid lines) and anions (dashed lines) for several representative shear rates, calculated from the EW simulations. The MSD curves for simulations with RF are plotted in Fig.~S4 in the supporting material. At shorter times, within the ballistic regime and over a portion of the plateau regime (until $\sim 3 \times  10^0$~[ns]), the MSDs remain unaffected by shear over the entire range of shear rates.
However, for larger times, towards the end of the plateau upon the onset of diffusive motion, the MSDs for distinct shear rates begin to diverge, resulting in higher self-diffusion with increasing shear rate.
At the same time, the differences between cation and anion self-diffusion slightly \emph{decrease} with increasing shear rate, and nearly vanish at the highest shear rate considered. 
Apparently, for high enough shear rates, subtle differences in specific interactions become less relevant, and the general structure and shape of the molecules, which define local interactions and arrangements, dominate the long-timescale motion.

This effect can be seen more clearly by examining the self-diffusion coefficient, determined by fitting $\langle \Delta y^2 \rangle(t) = 2D_yt$ within the diffusive regime of the MSD curves. Panel (b) of Fig.~\ref{fig:msd} presents the diffusion coefficients (in the $y$-direction) as a function of shear rate for both cations and anions and also for both EW (blue markers) and RF (red/orange markers) electrostatics.
In all cases, $D_y$ remains approximately constant at lower shear rates, before a rapid and systematic increase at some critical shear rate.
Cations and anions are affected at the same critical shear rate for a given electrostatic treatment. 
The EW simulations display a critical shear rate in the range $\dot\gamma \sim 10^{-1} $~ns$^{-1}$ to $\dot\gamma \sim 2 \times 10^{-1} $~ns$^{-1}$, consistent with the critical shear observed in both the structural and ISF analyses (Figs.~\ref{fig:rdf_peak}, \ref{fig:orientation_to_vector}, and \ref{fig:isf}).
The precise critical shear for RF simulations is less clear, although it appears similarly consistent within the range $\dot\gamma \sim 10^{-1}$~ns$^{-1}$ to $\dot\gamma \sim 3 \times 10^{-1}$~ns$^{-1}$. Despite the qualitatively similar behavior, the EW simulations exhibit diffusion constants within the low-shear regime that are about a factor of 2-3 higher than those in the corresponding RF simulations, consistent with the ISF analysis. Strikingly, this large difference quickly vanishes for the highest shear rates, further demonstrating the dominance of the shear flow over the system far from equilibrium.


\begin{figure}[ht]
    \begin{center}
        \includegraphics[width=0.9\linewidth]{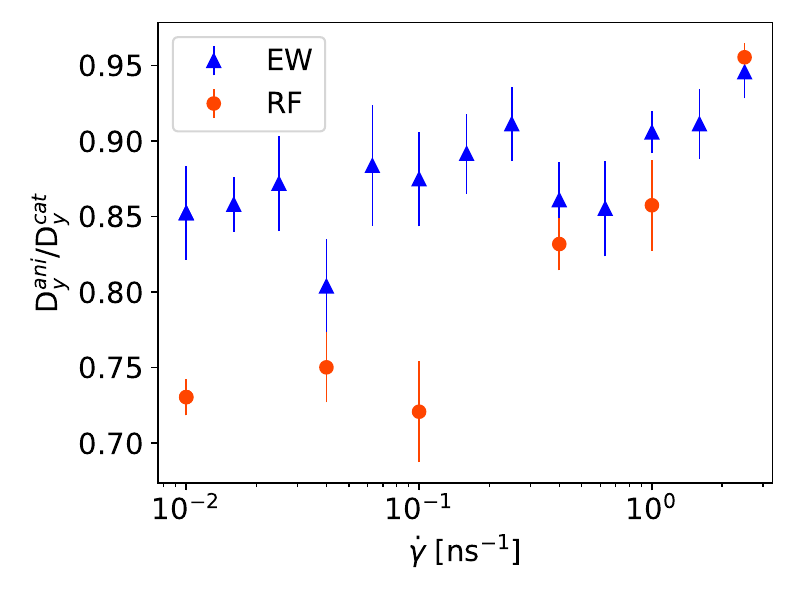}
        \caption{Shear rate dependence of the ratio between anion and cation self-diffusion coefficients compared between the two methods to calculate Coulomb interactions.}
        \label{fig:diffusion_comapre_anion_cation}
    \end{center}
\end{figure}

To quantitatively assess the relative change in the diffusion of anions and cations as a function of increasing shear, Fig.~\ref{fig:diffusion_comapre_anion_cation} presents the ratio of anion to cation diffusion constants, $D^{\rm ani}_{y}/D^{\rm cat}_{y}$, for both electrostatic treatments.
The cations are consistently faster in both models, similar to the reference atomistic model, as demonstrated in the previous investgation~\cite{Rudzinski_2021}.
Perfectly consistent with all the previous analysis, the RF ratio (red markers) shows a rather constant behavior until a critical shear rate in the range of $\dot\gamma \sim 10^{-1} $~ns$^{-1}$ to $\dot\gamma \sim 3 \times 10^{-1} $~ns$^{-1}$, beyond which the ratio approaches 1 with increasing shear. The EW ratio (blue markers) shows a similar trend, possibly with a slightly higher critical shear. The coincidentally higher ratio of EW at equilibrium reduces the magnitude of this trend, making it more subtle.

For a more detailed examination of the effect of shear on the IL dynamics, we calculate the non-Gaussian parameter $\alpha_2$, as a measure of the dynamic heterogeneity in the system.
As before, only the $y$-coordinates are taken into account and, thus, $\alpha_{2,y}$ can be expressed as:

\begin{equation}
    \alpha_{2,y}(t) = \left\langle \frac{\sum_{i=1}^{N} \left[y_i(t_0 + t)-y_i(t_0)\right]^4} {3 \left( \sum_{i=1}^{N} \left[y_i(t_0 + t)-y_i(t_0)\right]^2 \right)^2}  \right\rangle_{t_0} - 1 ,
    \label{eq:non_gaussian_parameter}
\end{equation}

where the variables are defined identically to Eq.~\ref{eq:msd} and the angle brackets represent an average over various time origins $t_0$. $\alpha_{2,y}$ is zero if the displacements of the particles from their initial position after a time $t$ follow exactly a Gaussian distribution and is positive if dynamical heterogeneities exist and lead to deviation from Gaussian statistics.

\begin{figure}[ht]
    \begin{center}
        \includegraphics[width=0.9\linewidth]{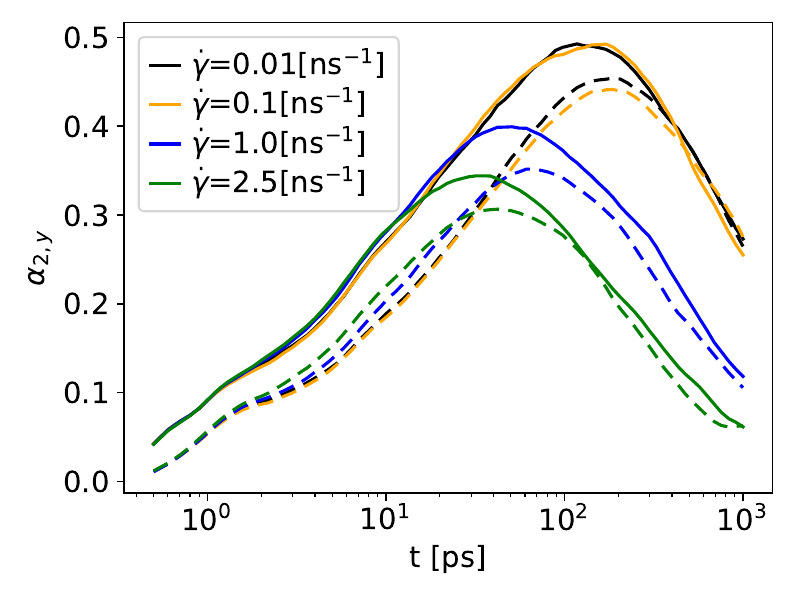}
        \caption{The non-Gaussian parameter in $y$-direction of the cations (solid lines) and anions (dashed lines) compared at different shear rates for the simulations with EW method.}
        \label{fig:non_gaussian_ew}
    \end{center}
\end{figure}

\begin{figure}[ht]
    \begin{center}
        \includegraphics[width=0.9\linewidth]{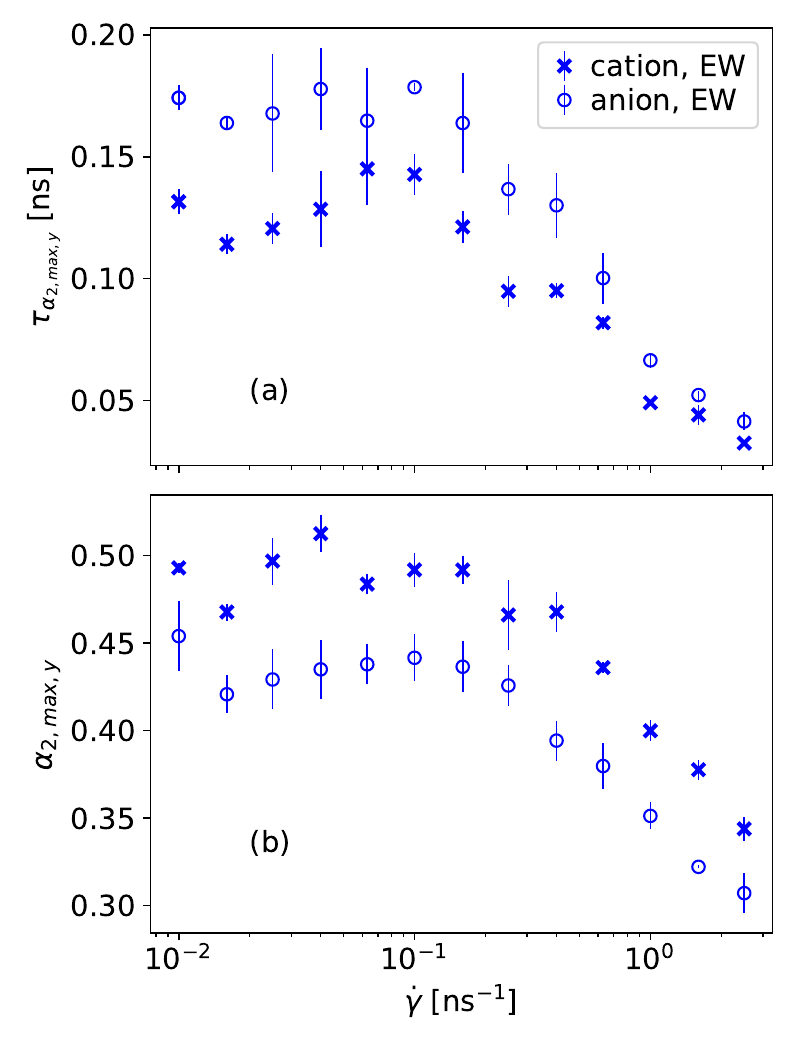}
        \caption{The maxima position (a) and their respective heights (b) from the non-Gaussian parameter for the different shear rates for the cations and anions. Only data from simulations with EW method are shown.}
        \label{fig:non_gaussian_max}
    \end{center}
\end{figure}

Fig.~\ref{fig:non_gaussian_ew} presents $\alpha_{2,y}$ for the cations (solid lines) and anions (dashed lines) for a set of representative shear rates. The results for the non-Gaussian parameter from the RF simulations are presented in Fig.~S5 in the supporting material.
Typical of viscous liquids~\cite{Henritzi2015-cm}, $\alpha_{2,y}$ shows a maximum at times corresponding to the onset of the structural relaxation. In particular, the maximum position of $\alpha_{2,y}$ for both cations and anions shifts to shorter times with increasing shear rate, in agreement with the acceleration of dynamics probed by ISF and MSD. Moreover, for a given shear rate, the maximum occurs at shorter times for cations than anions, consistent with the faster cation diffusion mentioned above. The maxima also decrease in height with increasing shear, indicating a reduced dynamical heterogeneity as the shear force begins to overpower the specific molecular interactions. These effects are more clearly seen in Fig.~\ref{fig:non_gaussian_max}, where the time corresponding to the maximum of the non-Gaussian parameter ($\tau_{\alpha_{2, max, y}}$) and its peak value ($\alpha_{2, max, y}$) are plotted as a function of the shear rate. In particular, it is evident from this figure that the onset of the shear rate dependence again occurs at a critical shear rate of $\dot\gamma \sim 10^{-1}\,$ns$^{-1}\,$.

\begin{figure}[ht]
    \begin{center}
        \includegraphics[width=0.9\linewidth]{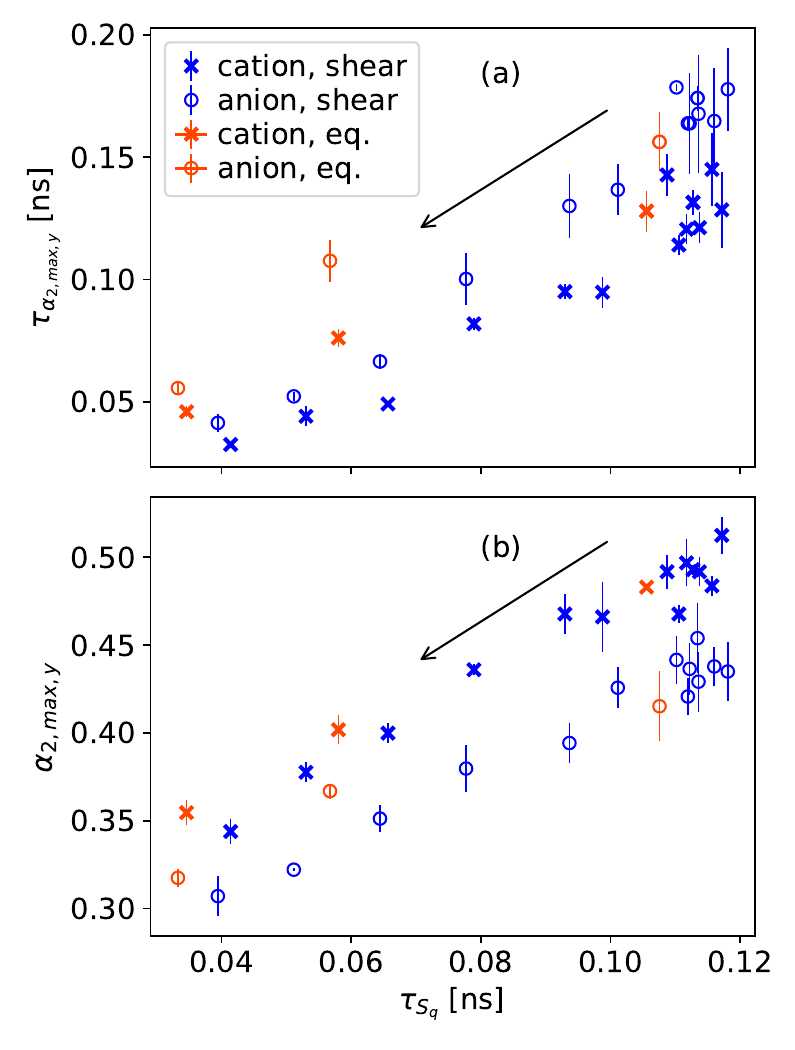}
        \caption{The maxima position (a) and their respective heights (b) from the non-Gaussian parameter compared to the correlation times from the ISF ( $\tau_{S_q}$) for the cations (cross markers) and anions (circles). Only data from simulations with EW method are shown. Additionally, data from equilibrium simulations between the temperatures of 300\,K, 320\,K, and 340\,K are added for comparison in red color. The arrows indicate the direction of increasing shear rate and temperature.}
        \label{fig:non_gaussian_max_vs_isf}
    \end{center}
\end{figure}

Previous simulations of viscous liquids have shown that deviations from Gaussian displacement statistics increase at lower temperatures, corresponding to longer structural relaxation times (\(\tau_{S_q}\)) \cite{PhysRevLett.96.057803, PhysRevE.60.3107}. This raises the question of whether the dependence of the height and position of the \(\alpha_{2,y}\) peak on shear rate is solely a consequence of changes in \(\tau_{S_q}\) or if there are additional shear-specific effects. 
To investigate this, Fig.~\ref{fig:non_gaussian_max_vs_isf} presents \(\tau_{\alpha_{2,\text{max},y}}\) and \(\alpha_{2,\text{max},y}\) as functions of \(\tau_{S_q}\) for both shear-flow simulations with various shear rates (blue markers) and also equilibrium simulations at various temperatures (red markers).
This result demonstrates that the dependence of the \(\alpha_{2,y}\) peak’s position and height on \(\tau_{S_q}\) is quite similar under both shear and equilibrium conditions, with even minor differences between anions (circle markers) and cations (x markers) remaining unchanged when varying \(\tau_{S_q}\) via either shear or temperature. This suggests that the reduction in dynamical heterogeneity with increasing shear is primarily driven by the acceleration of structural relaxation.

\section{Conclusions}\label{sec:conclusions}

In this study, we investigated the behavior of ionic liquids under shear flow using coarse-grained molecular dynamics simulations, with a focus on the impact of electrostatic treatments and the emergence of shear-induced structural and dynamical changes. By implementing the proper Lees-Edwards boundary conditions, while ensuring consistent treatment of both Ewald and reaction field electrostatics within this framework, we systematically assessed the influence of long-range interactions on shear-driven phenomena.

Our findings demonstrate that, while local structure appears to be only mildly perturbed by the less accurate electrostatics treatment, dynamical properties on all scales are significantly affected.  
These differences, however, diminish at higher shear rates, where the imposed shear dominates molecular motion and local arrangement. Importantly, we identified a critical shear rate, $\dot\gamma \sim 10^{-1}\,$ns$^{-1}\,$, beyond which both structural and dynamical properties deviate from equilibrium behavior. 
Furthermore, we observed a progressive reduction in dynamic heterogeneity with increasing shear. We demonstrated that this is not a shear-specific effect, but rather a consequence of the accelerated structural relaxation under flow. This trend parallels temperature-driven changes observed in equilibrium simulations, reinforcing the idea that shear-induced dynamics can be understood within a broader framework of relaxation processes.

Overall, our results provide key insights into the non-equilibrium behavior of ionic liquids, while highlighting the suitability of coarse-grained models for studying process-dependent material properties. Further insights may be gained from future investigations that directly compare with a compatible atomically-detailed model of this particular ionic liquid. In general, however, the methodologies applied in this study demonstrate promise for understanding shear-driven transport phenomena and rheological properties of ionic liquids, with direct relevance for lubrication, energy storage, and materials processing applications.

\begin{acknowledgments}

This work was supported by the TRR 146 Collaborative Research Center (Project A6) of the Deutsche Forschungsgemeinschaft. JFR was partially supported by the NFDI consortium FAIRmat - Deutsche Forschungsgemeinschaft (DFG) - Project 460197019.

\end{acknowledgments}

\section*{Data Availability}
\label{sec:data-avail}

As described in more detail in Section~\ref{sec:data-manag}, the data that supports the findings of this study are available on the NOMAD central repository within the dataset ``Coarse-grained shear-flow molecular dynamics simulations of the ionic liquid [C$_4$mim]$^+$ [PF$_6$]$^-$'' with DOI \url{https://dx.doi.org/10.17172/NOMAD/2025.04.15-1}.

\bibliographystyle{plain}
\bibliography{references}

\end{document}